**Error threshold estimation by means of the [[7,1,3]] CSS quantum code**

P. J. Salas[(1)], A.L. Sanz[(2)]

[(1)]Dpto. Tecnologías Especiales Aplicadas a la Telecomunicación,

[(2)]Dpto. Física Aplicada a las Tecnologías de la Información,

E.T.S.I. Telecomunicación, U.P.M., Ciudad Universitaria s/n, 28040 Madrid (Spain)

Short title: Error threshold estimation

ABSTRACT

The states needed in a quantum computation are extremely affected by decoherence. Several methods have been proposed to control error spreading. They use two main tools: fault-tolerant constructions and concatenated quantum error correcting codes. In this work, we estimate the threshold conditions necessary to make a long enough quantum computation. The [[7,1,3]] CSS quantum code, together with the Shor method to measure the error syndrome is used. No concatenation is included. The decoherence is introduced by means of the depolarizing channel error model, obtaining several thresholds from the numerical simulation. Regarding the maintenance of a qubit stabilized in the memory, the error probability must be smaller than 2.9 $10^{-5}$. In order to implement a one or two qubit encoded gate in an effective fault-tolerant way, it is possible to choose an adequate non-encoded noisy gate if the memory error probability is smaller than 1.3 $10^{-5}$. In addition, fulfilling this last condition permits us to assume a more efficient behaviour compared to the equivalent non-encoded process.

1. INTRODUCTION

It is believed that quantum computers will become a very powerful tool capable of performing calculations much faster than classical ones. The features providing this power are parallelism and interference, which are intrinsically quantum properties. They require creation and manipulation of entangled states involving large ensembles of quantum bits (qubits). Unfortunately quantum states are very fragile because their unavoidable interaction with the environment produces the qubit decoherence[1]. This process introduces errors into the computation, making it useless. To combat error accumulation, Shor[2] and Steane[3] introduced the concept of quantum error correction codes, capable of recovering some of the lost information once the qubit is suitably encoded. This possibility opened the door to a huge number of papers in the field[4,5]. Unfortunately, error-correcting methods are not strong enough to achieve a total control of error spreading through a quantum algorithm. In trying to solve this problem, Shor introduced fault-tolerant methods[6] in quantum computation. The basic idea is to apply quantum gates directly to the encoded qubits and to correct errors periodically. In addition, the encoded gates applied must be very carefully designed in order to restrict the error spreading towards the information. Roughly speaking, a fault-tolerant recovery method would introduce fewer errors than those it is able to eliminate. The implementation of a quantum algorithm requires a quantum computer to keep working on the qubits for a long time. The following step to improve the error control is the use of a concatenated quantum code[7] involving a hierarchical encoding structure. The fusion of fault-tolerant encoded quantum gates and concatenated codes has established the existence of an error threshold. If evolution and gate errors are below this threshold, quantum states will remain stabilized for a time long enough to carry out the computation. Several estimations for the value of this threshold have been published[8,5] using different error models and correction schemes. Gottesman and Preskill[9], by means of the same [[7,1,3]] code and concatenation estimate an error threshold rate of about $10^{-5}$ per time step, when memory error dominates. Following a method closer to the present one, Zalka[10] estimates the memory error threshold $\varepsilon$ and the gate error threshold $\gamma$ as one higher order of magnitude ($10^{-4}$) for the memory error. In an extensive[11] treatment using different encoding and recovery strategies, Steane finds an appreciably high threshold. Recently Reichardt[12] provides a smaller threshold ($9\ 10^{-3}$) than the present one, using the [[7,1,3]] quantum code and the depolarizing error model but without memory errors.

The aim of this work is to estimate the conditions to stabilize a qubit in memory to a great extend when no concatenation is used. We simulate numerically the stabilisation of a qubit exposed to memory (or free evolution) errors by means of a fault-tolerant recovery method (affected by the same memory errors as well as a gate error). The qubit is encoded using the [[7,1,3]] quantum code and the syndrome measurement is carried out applying Shor's method. Taking into account the results achieved, the one (Z gate) and two qubit (CNOT) gates threshold are estimated.

The structure of the paper is as follows. After a brief introduction, in section 2 we establish the main characteristics of the noise model and quantum code used. Starting from the well-known error equivalence in the code, in subsection 2.3, we provide a new and simple picture to describe how the different errors affect the quantum encoded qubit state. In section 3 we show the error correction procedure by means of the [[7,1,3]] quantum code. The results obtained in subsection 2.3 allow us to demonstrate how the fidelity depends on the error weight. As the fidelity is almost constant with the qubit considered, we use the simplest one ($|0_L\rangle$) in our numerical simulation. Finally, section 4 provides a new numerical estimation for the memory error as well as for the one and two qubit gate error threshold.

## 2. QUANTUM ERRORS

### 2.1 DISCRETIZATION AND ERROR MODEL

Quantum errors are continuous because they affect the coefficients in the qubit. The first step is to transform them into discrete errors[13].

The interaction of the initial information qubit (IQ in the following) $|q(0)\rangle = a|0\rangle + b|1\rangle$ with the environment produces the entanglement between both systems and, consequently, the qubit decoherence. Its time evolution may be written as a linear combination:

$$|q(0)\rangle \otimes |e\rangle \xrightarrow{\text{decoherence}} |q(t)\rangle = \{|e_I\rangle\hat{I} + |e_X\rangle\hat{X} + |e_Y\rangle\hat{Y} + |e_Z\rangle\hat{Z}\}|q(0)\rangle$$

where $\{|e_i\rangle, i=I,x,y,z\}$ are the environment states (neither normalised nor orthogonal). The operators $\{\hat{I}, \hat{X}, \hat{Y}, \hat{Z}\}$ are the Pauli matrices and represent the basic qubit discrete evolutions. We are allowed to interpret the qubit evolution as having one of these three errors ($\hat{I}$ is not actually an error): bit-flip, bit plus phase-flip, and phase-flip errors, respectively. This identification will only be strict when an orthogonal environment basis is used.

The classical error model (or channel) par excellence, considers the errors in different bits as independent. Even if this model does not adjust to the reality, it can provide some valuable consequences. In quantum information it is possible to introduce an analogous noisy channel called a depolarising error model. The environment states $\{|e_i\rangle, i=I,x,y,z\}$ are orthogonal and its scalar product is $|\langle e_i|e_j\rangle|^2 = \delta_{ij}\eta/3$ $(i,j\neq I)$, where $\eta/3$ is the probability (constant) of one of the three possible errors taking place, whereas the probability of no error is $|\langle e_I|e_I\rangle|^2 = (1-\eta)$. The qubit evolution can be represented by the operator $\hat{U}_D$:

$$\hat{U}_D(|q(0)\rangle \otimes |e\rangle) = \left\{\sqrt{(1-\eta)}|e_I\rangle\hat{I} + \sqrt{\frac{\eta}{3}}[|e_X\rangle\hat{X} + |e_Y\rangle\hat{Y} + |e_Z\rangle\hat{Z}]\right\}|q(0)\rangle$$

The error model is not unrealistic if one assumes that the physical qubits are located at sufficiently separated spatial positions, as in an ion-trap implementation of a quantum computer[14].

The qubit decoherence effect as well as the recovering (error correction) can be represented by superoperators[15]. If $\hat{D}^{(1)}$ is a superoperator or quantum operation representing the decoherence effect on the physical qubit and $\rho(0)$ is the initial density matrix, at time t the qubit is in a mixed state characterized by the final density matrix:

$$\rho_f(t) = \hat{D}^{(1)}(\rho(0)) = (1-\eta)\rho(0) + \frac{\eta}{3}\sum_{i=x,y,z}\hat{A}_i\rho(0)\hat{A}_i^+$$

The one-qubit error operators $\hat{A}_i \in \{\hat{X}(i=x), \hat{Y}(i=y), \hat{Z}(i=z)\}$.

Encoding by means of the [[7,1,3]] quantum code, the decoherence superoperator is represent by $\hat{D}^{(7)}$. If the recovery is perfect (without errors), the action of the $\hat{R}_{perfect}$ superoperator is:

$$\hat{R}_{perfect}\{\hat{D}^{(7)}(\rho(0))\} = \{(1-\eta)^7 + 7\eta(1-\eta)^6\}\rho(0) + O(\eta^2)$$

However, the real situation is more complicated because we will have to take into account that the recovery process itself is a quantum computation, so it is affected by decoherence. Writing an explicit equation for the full density matrix is complicated in this case. Instead of that, we will carry out a numerical simulation for the recovering process.

As much as we are interested in handling and transmitting quantum information just as if we consider the possibility of some type of encoding or quantum computation, we will have sets of n qubits called quantum registers $|q_1 q_2 ... q_n\rangle$. To see how the decoherence affects the registers, we can hypothesize on the error model to simplify the problem and constitute an approach to the reality[16]:

a) *Locally independent errors*.

If the environments to which the qubits interact (in the same time step) are different and not correlated, the errors in different qubits will be independent.

b) *Sequentially independent errors*.

The errors in the same qubit during different time steps are not correlated.

c) Assume a *small interaction qubit-environment.*

d) *Error-scalability independence*.

The qubit error probability is independent of the number of them that are used in the computation.

e) *No qubit leakage.*

The computation basis $\{|0>, |1>\}$ is sufficient to describe the qubit evolution; there is no qubit leakage outside this Hilbert subspace.

Under these hypotheses, errors that affect an increasing number of qubits are less probable and the operators representing the errors for an n-qubit register are the tensor product of those one-qubit operators[4]:

$$\hat{A}_{\{i_1, i_2, ..., i_n\}} = \hat{A}^1_{i_1} \otimes \hat{A}^2_{i_2} \otimes ... \otimes \hat{A}^n_{i_n}$$

where the superscript refers the qubit, and the subscript varies from 0 to 3: $\hat{A}^m_{i_m}$ (for the mth qubit) $\in \{\hat{I}\,(i_m=0),\, \hat{X}\,(i_m=1),\, \hat{Y}\,(i_m=2),\, \hat{Z}\,(i_m=3)\}$. In the depolarising error model, the evolution of an n-qubit quantum register is:

$$\hat{U}(|q_1 q_2 ... q_n\rangle |e\rangle) = |Q(t)\rangle =$$

$$= \left\{ (1-\eta)^{n/2} (\hat{I}_1 \otimes ... \otimes \hat{I}_n)|e_0\rangle + (1-\eta)^{(n-1)/2} \sqrt{\frac{\eta}{3}} \sum_{i=x,y,z} \left\{ \hat{A}_i \otimes \hat{I}_2 \otimes ... \otimes \hat{I}_n |e^1_i\rangle + \cdots + \right. \right.$$

$$\left. + \hat{I}_1 \otimes ... \otimes \hat{I}_{n-1} \otimes \hat{A}_i |e^n_i\rangle \right\} + ... + \left(\frac{\eta}{3}\right)^{n/2} \sum_{i_1, i_2, ..., i_n = x,y,z} (\hat{A}^1_{i_1} \otimes ... \otimes \hat{A}^n_{i_n}) |e^{1,...,n}_{i_1...i_n}\rangle \right\} |q_1 q_2 ... q_n\rangle$$

As the interaction with the environment is small (hypothesis c), the successive terms decrease quickly. A measurement of the register $|Q(t)>$ will produce a collapse in one of the terms according to its probability. Since each error $\hat{A}^k_{i_k}$ ($i_k \neq 0$) corresponds to three terms $\{\hat{X}, \hat{Y}, \hat{Z}\}$ and the probability of an error appearing in a given qubit is η, the one in which m errors appear in the register is P(m) = $\binom{n}{m}$ (1-η)$^{n-m}$η$^{m}$, describing a Bernouilli process of (1-η) probability. If η is small enough, the term with greater collapse probability is a register without error.

## 2.2 [[7,1,3]] QUANTUM CODE

To carry out the qubit decoherence control, we encode it by means of the [[7,1,3]] CSS (Calderbank-Shor-Steane) quantum code[3]. The quantum encoding protocol is based on the classical (linear) 7-bit Hamming code [7,4,3]. A 7-bit block is used to encode 4 bits of information ($2^4$ strings) and a 3-bit block ($2^3$ strings) to store the error information or syndrome. In the quantum code, the qubits are encoded by means of *cosets* and not directly with codewords as in classical encoding. The starting point is the dual even subcode $C^\perp \equiv [7,3,4]$ of the Hamming code $C \equiv [7,4,3]$, and the encoding uses the two *cosets* of C *relative* to $C^\perp$ (elements of the factor group $C/C^\perp$). The physical $|0\rangle$ is encoded as a logical $|0_L\rangle$ expressed as the linear combination of all codewords from the (0000000) coset (eight even parity Hamming codewords) and the physical $|1\rangle$ is encoded as a logical $|1_L\rangle$, the linear combination of all the (1111111) coset codewords, having odd parity:

$$|0_L\rangle = \frac{1}{\sqrt{8}} \left\{ \begin{array}{l} |0000000\rangle + |0001111\rangle + |0110011\rangle + |0111100\rangle + \\ |1010101\rangle + |1011010\rangle + |1100110\rangle + |1101001\rangle \end{array} \right\}$$

$$|1_L\rangle = \frac{1}{\sqrt{8}} \left\{ \begin{array}{l} |1111111\rangle + |1110000\rangle + |1001100\rangle + |1000011\rangle + \\ |0101010\rangle + |0100101\rangle + |0011001\rangle + |0010110\rangle \end{array} \right\}$$

## 2.3 ERROR EQUIVALENCE IN THE [[7,1,3]] CODE[17]

The relationship between the codewords of both classical codes (C and $C^\perp$) and the quantum states involved in the whole error correction process is shown in figure 1. The set of 7-bit classical codewords (ijk…) (with i, j, k,…∈ {0, 1}) form a 7 dimension binary linear space $F_2^{(7)}$ with 128 classical codewords. The $C^\perp \subset C$ even subcode, generates a partition of the $F_2^{(7)}$ in 16 cosets or $F_2^{(7)}/C^\perp$ elements, called $C^\perp \oplus (\mathbf{a})$, $(\mathbf{a}) \in F_2^{(7)}$ is a representative element of the coset. The elements of $F_2^{(7)}/C^\perp$ are shown in figure 1 inside full line boxes. Representing $(0000000) = (\mathbf{0})$ and $(1111111) = (\mathbf{1})$, the cosets can be expressed as $C^\perp \oplus X_\mathbf{v}(\mathbf{0})$ or $C^\perp \oplus X_\mathbf{v}(\mathbf{1})$, $\mathbf{v} = (v_1,…,v_7) \in F_2^{(7)}$ with weight $W(\mathbf{v}) = 0$ or 1. We can distinguish two special cosets (when $W(\mathbf{v}) = 0$, $\mathbf{v} = (\mathbf{0})$ and $X_\mathbf{0} = I$ (identity)) and their codewords will be directly involved in the quantum code: $C^\perp \oplus (\mathbf{0})$ (bottom of figure 1) and $C^\perp \oplus (1111111) \equiv C^\perp \oplus (\mathbf{1})$ (top of the figure 1). If $W(\mathbf{v}) = 1$, the $X_\mathbf{v}$ represent the bit-flip operator applied to the qubit where the component $v_i = 1$ (i=1,..,7).

In a similar way the code C induces a partition of the $F_2^{(7)}$ binary linear space in 8 cosets (translation of C by $F_2^{(7)}$) or elements of the factor group $F_2^{(7)}/C$, identified by means of a representative element $(\mathbf{b}) \in F_2^{(7)}$ ($W(\mathbf{b}) = 0$ or 1) as $C \oplus (\mathbf{b})$. As $C^\perp \subset C$, each coset $C \oplus (\mathbf{b})$ contains two cosets of $C^\perp$ such as $C \oplus (\mathbf{b}) = C^\perp \oplus X_\mathbf{b}(\mathbf{0}) \cup C^\perp \oplus X_\mathbf{b}(\mathbf{1})$. Note $C^\perp \oplus X_\mathbf{0}(\mathbf{0}) = C^\perp \oplus (\mathbf{0})$ and $C^\perp \oplus X_\mathbf{0}(\mathbf{1}) = C^\perp \oplus (\mathbf{1})$. The codewords of $C \oplus (\mathbf{b})$ cosets are shown inside 7 doted vertical boxes together an eighth $C^\perp \oplus (\mathbf{0}) \cup C^\perp \oplus (\mathbf{1})$ (top and bottom of figure 1). Bit-flip operators (errors) are represented as 7-component vectors $\mathbf{e}$, having 1's in each error location. The number of 1's in $\mathbf{e}$ is called the error weight $W(\mathbf{e})$. If a codeword $\mathbf{v}$ is affected by an error $\mathbf{e}$, the new codeword is $\mathbf{v} \oplus \mathbf{e}$ (mod 2) = $X_\mathbf{e}\mathbf{v}$ (mod 2). This fact allows us to follow the codeword paths in figure 1, when they are affected by errors.

The previous classical codeword distribution will help us to understand the effect of the different errors in the quantum CSS Q = [[7,1,3]] code. Each 7-bit classical codeword (ijk…) (with i, j, k,… ∈ {0, 1}) is transformed in a natural way to the quantum register (or state) |i j k…>. Loosely speaking, we will be able to consider quantum states obtained as a linear combination of $C^{\perp}$-coset classical codewords. Quantum states built in this way form a 128 dimension Hilbert space ($H_2^{(7)}$) basis and the quantum code Q is a two dimension subspace. From the 8 classical codewords of each $C^{\perp}$ coset, $C^{\perp} \oplus X_v(\mathbf{a})$ (**a** = **0**, **1**, W(**v**) = 1), we can build an 8-dimension subspace of $H_2^{(7)}$, noted as $[C^{\perp} \oplus X_v(\mathbf{a})]$ and characterized by a quantum orthonormal basis $\{X_v Z_w | a_L>\}$, W(**v**), W(**w**) = 0 meaning no error and W(**v**), W(**w**) = 1 represent the qubit affected by the error. This quantum meaning of each $C^{\perp}$ coset is also included in figure 1. The encoded |$0_L$> and |$1_L$> are in $[C^{\perp} \oplus (\mathbf{0})]$ and $[C^{\perp} \oplus (\mathbf{1})]$, respectively. The effect of an $X_v Z_w$ error with weight one (**v**, **w** ∈ $F_2^{(7)}$, W(**v**) = 1 or W(**w**) = 1), will convert the code Q = {|$0_L$>, |$1_L$>} in $X_v Z_w Q$ = {$X_v Z_w$|$0_L$>, $X_v Z_w$|$1_L$>} whose states are the linear combination of the classical codewords in the cosets $C^{\perp} \oplus X_v(\mathbf{0})$ and $C^{\perp} \oplus X_v(\mathbf{1})$, respectively.

A general information qubit (IQ) state |q(0)> will be encoded as |Q(0)$_E$> = a|$0_L$>+ b|$1_L$> and could be affected by an error during the computation or transmission process. As the [[7,1,3]] CSS code correct the $X_e$ and $Z_e$ errors independently, let's see its effect in the code states representing them as displacements in figure 1.

Bit-flip errors of weight one ($X_e$, W(**e**) = 1) transform the |$0_L$> (|$1_L$>) state, at the bottom (top) of the figure 1, into one of seven different orthogonal states, depending on the position of the '1' in the **e** vector and can be represented as $X_e$|$0_L$> ($X_e$|$1_L$>). As they are in two different cosets of $C^{\perp}$: $C^{\perp} \oplus X_e(\mathbf{0})$ and $C^{\perp} \oplus X_e(\mathbf{1})$, will have two different 4-bit syndromes ($s_1$, $s_2$, $s_3$; $s_4$) characterizing them. Fortunately as both quantum states are in

the *same* coset of C (C⊕(**e**)), the first three component ($s_1$, $s_2$, $s_3$) characterizing the syndrome according to the code C, will be the same. This fact will allow us to maintain the qubit coherence when correcting the encoded qubit. As the codewords of C have distance 3, the quantum code will be able to correct all errors of weight one.

Bit-flip errors of weight W(**e**) = 2 in different qubits (two errors in the same qubit are no error) produce $\binom{7}{2}$ = 21 different errors. As the classical codewords involved in the cosets $C^\perp \oplus (\mathbf{0})$ and $C^\perp \oplus (\mathbf{1})$ belong to C = [7,4,3] code, it is easy to check that for any error $\mathbf{e} \in F_2^{(7)}$ (W(**e**) = 2), there is a $\mathbf{v} \in F_2^{(7)}$, W(**v**) = 1 fulfilling $X_e|0_L\rangle = X_v|1_L\rangle$. Even though the errors are detectable, after carrying out the error correction, the encoded qubit has interchanged the quantum code basis states $|0_L\rangle \leftrightarrow |1_L\rangle$, the recovering network not being able to detect the error in subsequent time steps.

Less evident is the effect of the $\binom{7}{3}$ = 35 bit-flip errors of weight three ($X_e$, W(**e**) = 3). Let's define the sets $C^\perp \oplus (\mathbf{1}) - \{(\mathbf{1})\} = C^\perp \oplus (\mathbf{1}^*)$ and $C^\perp \oplus (\mathbf{0}) - \{(\mathbf{0})\} = C^\perp \oplus (\mathbf{0}^*)$. For the seven codewords $\mathbf{e} \in C^\perp \oplus (\mathbf{1}^*)$ with W(**e**) = 3, $X_e|0_L\rangle = |1_L\rangle$ and $X_e|1_L\rangle = |0_L\rangle$ is fulfilled. In this case, the error correction does not detect any error and there will be a basis inversion $|0_L\rangle \leftrightarrow |1_L\rangle$. The error can neither be detected nor corrected in subsequent time steps. Defining the support of a vector **v** (supp(**v**)) as its set of components different of 0, for the rest 28 error codewords **e** with W(**e**) = 3, such as supp(**e**) ⊂ supp(**v**, **v** ∈ $C^\perp \oplus (\mathbf{0}^*)$) (the inclusion comes from the fact that vectors in ($C^\perp \oplus \mathbf{0}^*$) have weight four), there is a $\mathbf{v} \in F_2^{(7)}$ with W(**v**) = 1, fulfilling $X_e\{|0_L\rangle, |1_L\rangle\} = X_v\{|0_L\rangle, |1_L\rangle\}$. Now, after the error recovery, the correct state is restored.

The rest of bit-flip errors $X_{e'}$ having W(**e'**) ≥ 4, are equivalent[18] to errors **e** (**e'** ~ **e**) with W(**e**) ≤ 3 so the effective error weight of **e'** is $W_{eff}(\mathbf{e'}) = W(\mathbf{e}) \leq 3$:

$$W_{eff}(\mathbf{e'}) = \underset{\mathbf{u} \in C^\perp}{\text{Min}} \{W(\mathbf{u} \oplus \mathbf{e'})\}$$

It is possible to make an analogous treatment for the phase-errors. The effect of phase-errors can be treated by means of its equivalence to bit-flip errors through the Hadamard transformation. This fact corresponds to changing the basis inside a subspace [C⊕(**b**)] (vertical dotted boxes in figure 1), from the computational $\{|0_L>, |1_L>\}$ to the dual $\{|0_L>\pm|1_L>\}$ basis.

Phase-flip errors $Z_\mathbf{e}$ with W(**e**) = 1, transform the code basis $\{|0_L>, |1_L>\}$ of [C⊕(**0**)] = [C$^\perp$⊕(**0**)] ∪ [C$^\perp$⊕(**1**)] into $Z_\mathbf{e}|0_L>$ and $Z_\mathbf{e}|1_L>$ (W(**e**) = 1), involving linear combinations with half of their signs being negative, so $Z_\mathbf{e}|0_L> \in$ [C$^\perp$⊕(**0**)] and $Z_\mathbf{e}|1_L> \in$ [C$^\perp$⊕(**1**)], fulfilling $<0_L|Z_\mathbf{e}|0_L> = <1_L|Z_\mathbf{e}|1_L> = <0_L|Z_\mathbf{e}|1_L> = 0$. Note the last condition is satisfied for every value of W(**e**), because the scalar product involves codewords (actually their quantum states) coming from classical orthogonal codes. These relationships are the same in the dual basis $\{H|0_L>, H|1_L>\}$ for bit-flip errors, allowing their complete correction by means of the equivalence Z = HXH.

For the 21 phase-flip errors of weight two ($Z_\mathbf{e}$, with W(**e**) = 2), there is a $\mathbf{v} \in F_2^{(7)}$ with W(**v**) = 1 so that $Z_\mathbf{e} \equiv Z_\mathbf{v}$ in the code [[7,1,3]]. In this case, it is not possible to correct the error. Suppose a phase-flip error $Z_\mathbf{e}\{|0_L>, |1_L>\}$ with W(**e**) = 2. The first step in the correction transform the states to the dual basis, $H\{Z_\mathbf{e}|0_L>\} = X_\mathbf{e}(|0_L>+|1_L>) \sim X_\mathbf{v}(|1_L>+|0_L>)$ with W(**v**) = 1. After $X_\mathbf{v}$ correction the right state ($|1_L>+|0_L>$) is recovered, ending with $|0_L>$ state after the Hadamard rotation to the computational basis. Unfortunately, the state $|1_L>$ is transformed into $Z_\mathbf{e}|1_L>$ and, after Hadamard rotation $H\{Z_\mathbf{e}|1_L>\} \sim X_\mathbf{v}(|0_L>-|1_L>)$ with W(**v**) = 1. Once correction is over, the state ($|1_L>-|0_L>$) is obtained and rotating it back provides the $-|1_L>$ state. The whole recovery for a general qubit a|0<sub>L</sub>> + b|1<sub>L</sub>> will provide the a|0<sub>L</sub>> - b|1<sub>L</sub>> state after the correction and introduces an undetectable erroneous relative phase which can be accumulated over time.

There are 35 phase-flip errors of weight three. Analogous to the bit-flip case, seven of them ($Z_e$, W(**e**) = 3 and **e** $\in$ $C^\perp \oplus (1^*)$) produce the transformation |0<sub>L</sub>> $\rightarrow$ |0<sub>L</sub>> and |1<sub>L</sub>> $\rightarrow$ -|1<sub>L</sub>>. The rest of them (28 errors) having supp(**e**) $\subset$ supp(**v**, **v** $\in$ $C^\perp \oplus (0^*)$), are equivalent to weight one phase-flip errors, therefore being correctable.

We could establish a similar error equivalence for Y errors just by taking advantage of the relationship Y=XZ. Note the [[7,1,3]] code can correct every error of weight one, bit-flip (X), phase-flip (Z) and both (Y) in the same qubit or both (bit-flip and phase flip) in *different* qubits.

## 3. ERROR CORRECTION WITH THE [[7,1,3]] CODE

In the following, we describe the steps involved in the encoding, fault-tolerant syndrome extraction and error correction as well as the measure of the qubit quality after the whole recovery process.

### 3.1 ENCODING

The C code generation matrix is used to implement the encoding network[3,19] appearing in figure 2. The initial IQ state $|q(0)\rangle$ is encoded as the quantum register $|Q(0)_E\rangle = a|0_L\rangle + b|1_L\rangle$ and could be affected by an error during the memory evolution process or computation (application of an encoded gate). If so, it must be corrected for the three kinds of error previously considered.

In order to recover the information, the error pattern is copied into the ancilla state and then this state is measured (collapsed into a particular codeword) to extract the error syndrome.

3.2 FAULT-TOLERANT ERROR CORRECTION: OBJECTIVES

Our goal is to maintain a qubit in the memory stabilized against noise. With this objective in mind, we have to take into account the errors introduced by the qubit recovery network, a quantum process itself. Therefore, the correction scheme and the individual gates must be implemented *fault tolerantly*[9]. In a quantum network, the total error probability introduced ($\eta$), originates in the evolution error (with probability $\varepsilon$ per qubit and time step) and gate error (with probability $\gamma$ for one-qubit gates and proportional to $\gamma$ for two qubit gates and per time step). The fault-tolerant network for a quantum code correcting one error will have an unrecoverable error probability behaving as $O(\varepsilon^2, \gamma^2)$. For a more general t-error correcting code, fault tolerance would requires that the error probability must behave as $O(\varepsilon^{t+1}, \gamma^{t+1})$.

The implementation of this criterion in the recovery involves several ideas: quantum gates applied directly on the encoded qubits (the present code permits to implement them *transversally*) in such a way that decoding is never required prior to the gate application and, finally, error correction is repeated periodically. Beside that, to build the appropriate recovery quantum networks we have to consider two facts: first, the recovery network is a quantum computation too, thus affected by errors and second, the gates used could propagate the errors towards the IQ. All these facts require a careful network design.

3.3 HIGH FIDELITY SHOR ANCILLA SYNTHESIS

The recovery uses an ancilla state to copy on the error syndrome, but a single error in the ancilla synthesis could propagate through CNOT gates in two or more unrecoverable errors, resulting in a fatal qubit contamination. So, high fidelity ancillas must be synthesized. Shor's method[6,20] proposes an ancilla state starting from a 4-qubit "cat" state (because four checks are needed to obtain each bit of syndrome):

$$|\text{cat}\rangle = \frac{1}{\sqrt{2}}\left(|0000\rangle + |1111\rangle\right)$$

which is four times Hadamard rotated to get Shor's ancilla state:

$$|\alpha_{\text{Shor}}\rangle = \frac{1}{\sqrt{8}}\begin{pmatrix} |0000\rangle + |0011\rangle + |0101\rangle + |0110\rangle \\ + |1001\rangle + |1010\rangle + |1100\rangle + |1111\rangle \end{pmatrix}$$

The detailed network is sketched in figure 3. It includes a fifth qubit to detect a possible bit-flip error. This error would become a phase-flip error through bitwise Hadamard gates at the end, and would spread back to the IQ, damaging it. These errors may be controlled applying two CNOT gates from the first and fourth qubits (as control) to the fifth qubit (as target). If the measurement of this last qubit is "1", the ancilla is wrong, is then rejected, and a fresh one is synthesized.

In spite of all that, some errors could slip through this verification if they happen at the end of the ancilla network, during or after the last few CNOT gates, but they are uncorrelated. Therefore, we can conclude that the ancilla synthesis can be done fault-tolerantly only against bit-flip errors, i.e. with an unrecoverable error probability behaving as $O(\varepsilon^2, \gamma^2)$.

Fault tolerance against phase-flip errors is achieved repeating the syndrome three times and making the qubit correction by means of the majority-vote method.

3.4 SYNDROME MEASURING AND ENCODED QUBIT CORRECTION

The next step copies the error syndrome onto the ancilla state without destroying the IQ coherence. CNOT gates and H gates are implemented in a fault-tolerant way applying them transversally[21].

To measure each bit of syndrome (see figure 4), we use a cat or ancilla state $|\alpha_{Shor}\rangle$. Each ancilla is a linear combination of eight even parity four qubit codewords. If the system has only one bit-flip error, the CNOT gates between the IQ and the ancilla, will change the parity in the ancilla codewords. Measuring this parity, the ancilla vector collapses into a particular codeword that provides one bit of syndrome. The whole qubit recovery takes 24 CNOT gates for one syndrome measurement. In the case of three-syndrome repetition, the network includes 72 CNOT gates.

Errors appearing in the middle of the qubit recovering process or coming from an erroneous ancilla synthesis behave as $O(\varepsilon,\gamma)$ and will contaminate the ancilla producing a wrong syndrome. If this syndrome were used to correct the IQ state, the error would be propagated into the information. To control this possibility, the syndrome measurement is repeated three times, choosing the correction action by a majority-vote method. This repetition decreases the error probability to $O(\varepsilon^2,\gamma^2)$ and makes the whole recovery method fault-tolerant.

It is easy to understand how the network used to detect bit and phase flips is built if we realise how CNOT gates transmit the errors[9]. For phase errors, we prepare a cat state in the computational basis, and then through some transversal CNOT gates we copy the possible phase-flip errors in the IQ to the cat state using their back propagation target-to-control. This fact can be understood easily considering the action of a CNOT gate connecting the control qubit $[a|0> + b |1>]_C$ and a target qubit $[|0> + |1>]_T$ affected by a phase error $\hat{Z}$:

$$^C\hat{X}\{[a|0\rangle + b|1\rangle]_C \otimes \hat{Z}[|0\rangle + |1\rangle]_T\} = \hat{Z}[a|0\rangle + b|1\rangle]_C \otimes \hat{Z}[|0\rangle + |1\rangle]_T$$

After the CNOT gate, the phase error in the target qubit has been propagated to the control qubit. Then phase errors are rotated and transformed into bit-flip errors that will be read out as the syndrome. This is indicated as phase-flip syndrome process in figure 4.

In the bit-flip error detection, the rotation has to be made before the CNOT gates in order to get the syndrome bits. The cat state is transformed into the full Shor ancilla state. This is shown in the three bottom lines of figure 4, labelled bit-flip syndrome.

The parity measurement of the ancillas provides the six bits of syndrome, three for phase-flips and three for bit-flips. Once the error syndrome is obtained, the qubit error may be identified and we are able to correct it.

## 3.5 RECOVERED QUBIT QUALITY

To test how good our whole process has been, when the recovery is over, we calculate the overlapping squared ($F(\varepsilon,\gamma,t,a)$) between the final corrected state and the error free initial state. This overlapping depends on the memory ($\varepsilon$) and gate ($\gamma$) error probability, time step t as well as the concrete qubit chosen $|q(0)> = a |0> + b |1>$ through the a and b coefficients. The fidelity is defined at the time step t, as the minimum value of $F(\varepsilon,\gamma,t,a)$, once the error correction has been carried out:

$$F(\varepsilon,\gamma,t) = \underset{\forall a\in[0,1]}{\text{Min}} F(\varepsilon,\gamma,t,a) = \underset{\forall a\in[0,1]}{\text{Min}} \langle Q(0)_E |\rho_f(t)| Q(0)_E \rangle = \underset{\forall a\in[0,1]}{\text{Min}} \left|\langle Q(t)_{E,\text{Corrected}} | Q(0)_E \rangle\right|^2$$

where $\rho_f(t) = |Q(t)_{E,\text{Corrected}}\rangle\langle Q(t)_{E,\text{Corrected}}|$ is the final encoded density matrix at time t. In a general process, its value depends on the memory and gate error as well as the time step considered.

The overall recovering process can be represented as a black box in which the initial state $|Q(0)_E\rangle$ is introduced at one end and we get the $|Q(t)_{E,\text{Corrected}}\rangle$ at time t, at the other end. The $|Q(t)_{E,\text{Corrected}}\rangle$ could contain some errors $X_e$ (bit-flips) and $Z_e$ (phase-flips) of effective weight $0 \leq W_{\text{eff}}(e) \leq 3$, with probabilities $\eta_i^b$ and $\eta_i^p$, being $i = W_{\text{eff}}(e)$. In terms of these probabilities $\{\eta_i^{b,p}\}$, the overlap can be written as:

$$F(\varepsilon,\gamma,t,a) = \langle Q(0)_E |\rho_f(t)| Q(0)_E \rangle = \eta_0 f_0 + \sum_{i=1}^{3}\left(\eta_i^b f_i^b + \eta_i^p f_i^p\right)$$

where $f_i^b = |\langle Q(0)_E|\hat{X}_e|Q(0)_E\rangle|^2$ and $f_i^p = |\langle Q(0)_E|\hat{Z}_e|Q(0)_E\rangle|^2$. The $\eta_0$ is the probability that the final state will be error free, so $f_0 = |\langle Q(0)_E|\hat{X}_0|Q(0)_E\rangle|^2 = 1$. If $W_{\text{eff}}(e) \leq 2$ then $f_{1,2}^{b,p} = 0$. For $W_{\text{eff}}(e) = 3$ two possibilities came out. If $\text{supp}(e) \subset \text{supp}(v, v \in C^\perp \oplus (0^*))$, $W_{\text{eff}}(e) = 1$ and $f_3^b = f_3^p = 0$. If $e \in (C^\perp \oplus (1^*))$, then $f_3^b = |2ab|^2$ and $f_3^p = |a^2-b^2|^2$. The overlap (squared) can be expressed as:

$$F(\varepsilon,\gamma,t,a) = \eta_0(t) + \eta_3^b(t)|2ab|^2 + \eta_3^p(t)|a^2-b^2|^2 = \eta_0(t) + \eta_3^p(t) + 4a^2(1-a^2)\Delta\eta_3(t)$$

where $\Delta\eta_3 = \eta_3^b - \eta_3^p$ is the difference between the final probability of bit-flip and phase-flip errors. To calculate the strict fidelity $F(\varepsilon,\gamma,t)$, we minimize $F(\varepsilon,\gamma,t,a)$ with respect to the a coefficient, getting the values a = 0 and a = b =$1/2^{1/2}$. For the former value $F(\varepsilon,\gamma,t)$ = $\eta_0(t) + \eta_3^p(t)$ and for the latter $F(\varepsilon,\gamma,t) = \eta_0 + \eta_3^p + \Delta\eta_3$, so the existence of a maximum or a minimum in a = b =$1/2^{1/2}$ depends on the $\Delta\eta_3$ sign, and hence of the whole recovery network. If the recovery network is well balanced, we expect $\Delta\eta_3 \sim 0$, and the fidelity would not longer depend on the particular qubit.

To check the dependence with the a coefficient, we have carried out a calculation for the fidelity after one recovery step (t = 20, final time step after applying three times the recovery network of figure 4, plus one time step to correct the information qubit), $F(\varepsilon,\gamma,t=20,a)$ versus a. The reference computation process used is a noisy encoding network shown in figure 2, according the error model that will be explained in section 4.1. The numerical simulation concludes that for the range considered $0.02 < \varepsilon, \gamma < 0.0002$, the values are $|\Delta\eta_3| \sim 10^{-2}$-$10^{-4}$, so the fidelity dependence with the coefficients is very weak. In figure 5 the fidelity, together with $\eta_3^{b,p}$ and $\Delta\eta_3$ are represented as a function of the a coefficient for $\varepsilon = 0.002$ and $\gamma = 0.02$. The $\eta_3^{b,p}$ probabilities and $\Delta\eta_3$, are independent of the qubit considered, but they depend on the $\varepsilon$ and $\gamma$, having positive or negative values in the interval studied. Therefore, for some $\varepsilon$ and $\gamma$, the minimum will be at a = 0 (or a = 1) and for some other $\varepsilon$ and $\gamma$ will be at a = b =$1/2^{1/2}$. In the case of figure 5, $\Delta\eta_3 < 0$ and the minimum of the fidelity appears at $|a|^2 = 0.5$. For the rest of the paper, and for simplicity we will use the $|0_L\rangle$ as the reference state to carry out the numerical simulation.

## 4. MEMORY AND GATE THRESHOLD ESTIMATION

### 4.1 DECOHERENCE MODEL AND NUMERICAL SIMULATION

To simulate the noisy quantum network involving quantum registers, the *independent stochastic error model* based on the notion of error locations[22] is used. In a given location of the network, a random error is introduced. Each error is independent of the other errors happening at the same or different locations. All quantum steps have some error probability, and we distinguish between *memory errors* (caused by qubit free evolution) with error probability $\varepsilon$, *one-qubit gate error*, as a result of one-qubit gate operation (like measurements or Hadamard gates) with $\gamma$ error probability and *two qubit gates* as the CNOT, with an error probability proportional to $\gamma$. The intrinsic $\gamma$ error probability affecting the qubits involved in the gate application is not the only source of gate noise. In addition, we consider that the implementation of a gate is carried out in one time step, so we include an additional evolution error (with $\varepsilon$ error probability) for all the physical qubits involved or not in the gate application. Therefore, the total qubit error in the gate time step operation is $\varepsilon$ for qubits not affected by the gate and $\varepsilon + O(\gamma)$ if they are.

Memory errors are located at each time step in the network, affecting all the qubits evolving in that step. For each error location affecting one physical qubit with an $\varepsilon$ error, we consider an isotropic $\varepsilon/3$ error probability for each of the $\sigma_X$, $\sigma_Y$ and $\sigma_Z$ operators. Their effect is highly related to the degree of parallelization, so one way to reduce this error is to increase the network parallelism. The error correction network takes into account the gate parallelization applying them in one time step, when they commute (gates affecting different qubits). For the noisy one-qubit gates (Hadamard and measurement), the $\gamma$ intrinsic error probability is introduced at each gate location together an $\varepsilon$ memory error probability. In the two-qubit gates (CNOT), we assume there are sixteen possibilities corresponding to the tensor product $\{I, \sigma_X, \sigma_Y, \sigma_Z\} \otimes \{I, \sigma_X, \sigma_Y, \sigma_Z\}$. If the intrinsic one-qubit gate error probability is $\gamma$, each two-qubit error appears with probability $\gamma/15$, because the $I \otimes I$ term is not, actually, an error operation. In all cases, we let the gate operate before the error is introduced. This $O(\gamma)$ (instead of $O(\gamma^2)$) two-qubit error behaviour clearly over-estimates the difficulty of error correction, although it is not an unrealistic assumption. Finally we assume that $\varepsilon$ and $\gamma$ errors are independent of the total number of qubits in the network. Introducing errors in this way is equivalent to collapse the qubit register $|Q(t)_E\rangle$ stochastically in one of its error terms.

The error type in each location and time step is chosen invoking a Monte Carlo simulation by means of the Luxury Pseudorandom Numbers[23] that is an improvement in the subtract-and-borrow random number generator proposed by Marsaglia and Zaman. The fortran-77 code is due to James[24], and is used with the highest luxury level parameter $p = 389$. As the code state for this value of p, any theoretically possible correlations have very small chance of being observed. The code returns a 32-bit random floating-point number in the range (0, 1). For each run, a new random seed is chosen as a 32-bit integer. In the numerical simulation, for each $\varepsilon$, $\gamma$ and time step, we carry out the whole error correction process a number of times $N \gg \max(1/\varepsilon, 1/\gamma)$, obtaining the fidelity as its average value.

4.2 MEMORY ERROR THRESHOLD ESTIMATION

The first objective is to preserve a qubit stored in the memory of a quantum computer when it is affected by an evolution error of probability $\varepsilon$. We should ask the following question: is it useful to encode and correct the qubit in order to stabilize it in the memory? Moreover, how big could be the gate error probability (affecting only the recovery network) to preserve the memory stabilization? To find an answer, in this work we have simulated the time evolution of the simplest qubit |0> encoded as $|0_L\rangle$, sent through a quantum channel affected only by a memory error of probability $\varepsilon$. The behaviour of the

'naked' qubit (neither encoded nor corrected) and the encoded qubit with the previous [[7,1,3]] quantum code, is compared.

The naked qubit fidelity follows the law $(1-2\varepsilon/3)^t$, because only two (X and Y) from the three possible errors produce a zero fidelity state (the Z error leaves the qubit state unchanged). For small enough $\varepsilon$ error, this fidelity behaves like the usual exponential decreasing law $\exp(-2\varepsilon t/3)$ and for not for too long, it can be approximated by a straight line with $-2\varepsilon/3$ slope.

The second possibility to send the qubit through the channel involves encoding the qubit according to the network shown in figure 2, send it through a quantum channel affected by memory errors of probability $\varepsilon$, and make a complete error recovery (by means of a network with the same evolution error $\varepsilon$ as well as a $\gamma$ gate error probability) at different times.

The effectiveness of the error recovery is studied in order to achieve two objectives. The first intention is to attain an uncorrectable error probability smaller than $1-(1-2\varepsilon/3)^t$. We imagine this will not be difficult because the error correction is fault-tolerant. The second goal will be to keep the qubit more stabilized during the time than the naked one. This is not evident because in both cases (encoded qubit or not) errors of weight one with probability $O(\varepsilon)$ will appear. The recovery network is capable of correcting some errors but, in addition, it introduces some noise into the encoded qubit. The balance between both processes depends on the network complexity and the error probability. Too many time steps and gates would not make the recovery useful even if the error probability were small. If the recovery network has a nice parallelization, we guess its ability to correct the errors would depend on the relationship between the evolution and gate error probability. If the gate error likelihood is too big, the recovery would be useless and if it is small enough, we hope to have a successful qubit correction. In the present simulation we have considered memory errors ($\varepsilon$) and intrinsic gate errors ($\gamma$) fulfilling the relationship $\varepsilon = C\gamma$, with $0.3 \leq C \leq 2$ and the case $\gamma \to 0$ implying $C \to \infty$. This latter case corresponds to a gate with only $\varepsilon$ evolution error.

In the naked qubit case (non-encoded), the uncorrectable error probability after t time steps is $P_{NE}(\varepsilon,t) = 1-(1-2\varepsilon/3)^t$, whose behaviour for $\varepsilon$ sufficiently small, is linear: $P_{NE}(\varepsilon,t) \sim 2\varepsilon t/3$. If an error occurs, there is no possibility of correcting the state to recover the initial qubit. When encoding is used, an error affecting one physical qubit at time $t_i$ is not very harmful, because it could be corrected at time $t_i+1$, by the method explained in the first part of this paper. Otherwise, as the error correcting method is fault-tolerant, the uncorrectable errors (of weight two o more) do not accumulate very quickly, and the uncorrectable error probability will behaves as $P_E \sim O(\varepsilon^2,\gamma^2)$ (see figure 6). When the encoded qubit reaches the receiver, the information recovering process will be largely successful if the final decoding step could be performed without any error. The receiver fidelity will behave as $1-O(\varepsilon^2,\gamma^2)$. So, for $\varepsilon = C\gamma$ small enough, there must be an $\varepsilon$-region in which $P_E < P_{NE}$.

We study (for each C) the $\varepsilon$ threshold for which the condition $P_E \leq P_{NE}$ is fulfilled. In the case of an encoded qubit, the first error correction is carried out at t = 20 time steps, so in figure 6 we represent the probabilities $P_{NE}(\varepsilon,t=20)$ and $P_E(\varepsilon,C,t=20)$. The curves $P_E(\varepsilon,C,t=20)$ are satisfactorily fitted to a quadratic polynomial $D_2\, \varepsilon^2$ (see table I), reflecting the fault-tolerance of the method. The crossing points between the curves $P_{NE}(\varepsilon,t=20)$ and $P_E(\varepsilon,C,t=20)$, provide the unrecoverable error probability-threshold $\varepsilon_{pth}(C) \sim 40/3D_2$ which are represented in figure 7. The region under the curve $\varepsilon_{pth}(C)$ is where the uncorrectable error probability for the encoded qubit is smaller than for the naked one.

Even though the encoded qubit is stabilized in its uncorrectable error probability, stabilization is not guaranteed over time. The second objective is, restricted to the uncorrectable error probability region determined above, to study the qubit stabilization over time.

To see the channel effect, the initial qubit is error free encoded and sent through the quantum channel, and its fidelity $F(\varepsilon,C,t)$ calculated after each recovery step. Two consecutive recoveries are separated by only one time step, $\Delta t_0 = \Delta t = 1$ (see figure 8).

Not all errors of weight one are eliminated because some of them appear at the end of the network so, as the numerical simulation shows, if $\varepsilon$ is small enough, the fidelities $F(\varepsilon,C,t)$ are perfectly fitted to straight lines of negative slope $F(\varepsilon,C,t) = - A(\varepsilon,C) \, t + B(\varepsilon,C)$. As the error probability $\varepsilon$ decreases, the slope $A(\varepsilon,C)$ tends towards zero. This means that, for $\varepsilon$ small enough, the recovery method has largely stabilized the qubit in the memory (or in the channel) of the quantum computer, i.e. fidelity remains almost constant as a function of time. In a certain sense, $A(\varepsilon,C)$ quantifies the stabilization degree of the qubit through the time. The next step will be to study the variation of the fidelity slope $A(\varepsilon,C)$ as a function of $\varepsilon$ for different $\varepsilon/\gamma = C$ ratios.

In figure 9, the slope $A(\varepsilon,C)$ is represented versus $\varepsilon$ for different $\varepsilon/\gamma = C$ values. We use the $A(\varepsilon,C) \leq 2\varepsilon/3$ condition to decide whether the encoding and correction is advantageous. These values of $\varepsilon$ determine the region in which the use of an encoded and corrected qubit is more stabilized than without encoding and error correction. In order to establish the crossing points, the numerical results for $A(\varepsilon,C)$ have been fitted to polynomials of order 2 and 3 in $\varepsilon$. The curves tend towards zero fidelity slope when the error $\varepsilon$ approaches 0, as expected, because there would be no error at all. Figure 9 provides the slope-threshold $\varepsilon_{sth}(C)$ (crossing points between $A(\varepsilon,C)$ and the $2\varepsilon/3$ line) that is represented in figure 7 versus C. The slope-stabilization region is under the curve. The horizontal (dashed) line above the curve, represent the value $\varepsilon_{sth}(C\to\infty)$, i.e. when $\gamma = 0$ and gates have only an $\varepsilon$ error.

Strictly speaking there is no absolute threshold value for $\varepsilon$ achieving a complete stabilisation ($A(\varepsilon,C) = 0$) of the qubit in a quantum memory (or throughout a quantum channel), because there will always appear uncorrectable errors (of weight two or more) with a finite probability.

In addition to the regions under the curves $\varepsilon_{pth}(C)$ and $\varepsilon_{sth}(C)$, two practical threshold values can be inferred from figure 7. Encoding the qubit with the objective to have an uncorrectable error probability smaller than the obtained with a naked qubit, requires $\varepsilon < \varepsilon_{pth}(C\to\infty)$ to be satified. Fulfilling this condition, assures the existence of a noisy gate with an error probability $0 < \gamma = \varepsilon/C$ (for the gates used in the recovery network) able to produce a higher quality qubit state than without encoding. So we propose to consider the value $\varepsilon_{pth}(C\to\infty) = 3.9\ 10^{-4}$ as a type of memory error threshold through the channel. Moreover, if we are looking for a threshold related to keeping the qubit more stabilized in memory than without encoding, we require (in the best case, i.e. without gate error, $\gamma = 0$) to satisfy the condition $\varepsilon \leq \varepsilon_{sth}(C\to\infty) = 2.5\ 10^{-4}$. This value assures that it is possible to keep the qubit stabilized in the quantum memory by means of a correcting network built using noisy gates with an error $0 < \gamma = \varepsilon/C$.

In addition to the previous values, a more severe threshold can be inferred requiring an effective fault-tolerant behaviour of the recovery network, i.e. the uncorrectable error probability should behave as $O(\varepsilon^2)$. Because $P_E(\varepsilon,C,t=20) = D_2\ \varepsilon^2$, we adopt as memory threshold region the one under the curve $\varepsilon_{mth}(C) = 1/D_2$, and its dependence on C is shown in figure 7. As the final memory threshold value we propose the value $\varepsilon_{mth}(C\to\infty)$ = $1/D_2(\infty) = 2.9\ 10^{-5}$.

Considering an encoded qubit sent through a quantum channel affected by a memory error probability $\varepsilon$, if the condition $\varepsilon < \varepsilon_{mth}(C\to\infty)$ is fulfilled, we can infer that there is a set of noisy gates (including H, $^C$X, and measurements) with an intrinsic error probability $0 < \gamma = \varepsilon/C$, able to construct a noisy fault-tolerant correcting network stabilizing the qubit in the quantum memory.

4.3 ONE-QUBIT GATE ERROR THRESHOLD CALCULATION

Using the same decoherence model, we have calculated the one-qubit gate threshold by means of the encoded fault-tolerant Z-gate. Applying a single error free Z gate to the naked qubit |0>, does not change the state. The total gate error probability comes from the intrinsic gate application (with $\gamma$ error) and one time step (with an $\varepsilon$ error probability) in which the gate is carried out (see figure 10), together with another time step that we consider previous to the gate (t step in figure 10). The total error probability for the non-encoded one-qubit gate application is $P_{NEg}^{(1)}(\varepsilon,\gamma) = 2/3(2\varepsilon+\gamma) = (2\varepsilon/3)(2+1/C)$, because only the X and Y (not the Z) operators produce an uncorrectable error.

When encoding is used, the Z gate can be implemented in a fault-tolerant way as a transversal gate applied bit wise (see figure 10) followed by a fault tolerant error correction step, at the end of which the probability for the appearance of uncorrectable errors, $P_{Eg}^{(1)}(\varepsilon,\gamma)$, is calculated. Assuming the relationship $\varepsilon = C\gamma$, we can express this probability as a function of $\varepsilon$ for each constant C as $P_{Eg}^{(1)}(\varepsilon,C)$. Following a similar method to the used in the previous paragraph, a gate threshold can be inferred from the condition $P_{Eg}^{(1)}(\varepsilon,C) \leq P_{NEg}^{(1)}(\varepsilon)$.

Considering the implementation model shown in the figure 10, the $P_{Eg}^{(1)}(\varepsilon,C)$ probability can be written explicitly. Several terms contribute to it:

a) Two error probability in each step:

$\binom{7}{2} (2\varepsilon/3)^2$ in the evolution time step t and in the time step associated to the Z-gate, $t_Z$

$\binom{7}{2} (2\gamma/3)^2$ in the transversal Z gate, and

two-error probability in the error correction step, $P_{EC_2}(\varepsilon,C)$

b) Crossed terms coming from one error in each of the steps t, $t_Z$, Z-gate and one error probability in the error correction step $P_{EC_1}(\varepsilon,C)$. The result is 196/9 ($\varepsilon^2$ +2 $\varepsilon\gamma$) + 14/3 (2$\varepsilon$ + $\gamma$) $P_{EC_1}(\varepsilon,C)$ = 196$\varepsilon^2$/9 (1+2/C) + 14$\varepsilon$/3 (2+1/C) $P_{EC_1}(\varepsilon,C)$

Taking into account that $P_{EC_2}(\varepsilon,C) = P_E(\varepsilon,C,t=20) = D_2(C) \varepsilon^2$, we only need to calculate the probability for one error in the correction process, $P_{EC_1}(\varepsilon,C) = D_1(C) \varepsilon$. The values for $D_1(C)$ are shown in table I and have been obtained fitting the simulation results for $P_{EC_1}(\varepsilon,C)$ to straight lines. The final expression for $P_{Eg}^{(1)}(\varepsilon,C)$ is:

$$P_{Eg}^{(1)}(\varepsilon,C) = \left\{ \frac{4}{9}\left(91 + \frac{98}{C} + \frac{21}{C^2}\right) + \frac{14}{3}\left(2 + \frac{1}{C}\right)D_1 + D_2 \right\} \varepsilon^2 = G_1 \varepsilon^2$$

The values for $G_1$ appear in table I. Now the condition $P_{Eg}^{(1)}(\varepsilon,C) \leq P_{NEg}^{(1)}(\varepsilon,C)$ provides a first one-qubit gate threshold as $\varepsilon_{g_1}(C) = 2(2+1/C)/3G_1$ (see figure 7). Surprisingly its value is almost constant in the range of C studied, so $\varepsilon_{g_1}(C)$ does not reflect a clear advantage of the encoded gate version vs. the non-encoded gate when C increases.

Analogous to the memory threshold case, we can obtain a more appropriate threshold value requiring its fault-tolerant behaviour. The $P_{Eg}^{(1)}(\varepsilon,C) = G_1\varepsilon^2$ and the proper threshold curve will be $\varepsilon_{thg_1}(C) = 1/G_1$, proposing the value $\varepsilon_{thg_1}(C \to \infty) = 2.7 \ 10^{-5}$ as the effective one-qubit gate threshold.

Given a non-encoded gate, if the memory error probability is $\varepsilon$, and $\varepsilon < \varepsilon_{thg_1}(C \to \infty)$, we can conjecture that the transversal (fault-tolerant) encoded one-qubit gate version using the [[7,1,3]] code, is advantageous with respect to the non-encoded gate. Its helpfulness concerns two aspects: first, it provides a smaller uncorrectable error probability than the non-encoded gate and secondly, it affords a fault-tolerant behaviour controlling the error spreading and accumulation.

4.4 TWO-QUBIT GATE ERROR THRESHOLD ESTIMATION

By means of the previous treatment, we can estimate the error probability for a typical two-qubit gate as the $^CX$. Without detailing all the error sources, the uncorrectable error probability can be approximated as $P_{Eg}^{(2)}(\varepsilon,C) = 2G_1\varepsilon^2$, considering an error probability of $G_1\varepsilon^2$ in each quantum register involved in the CNOT. The threshold curve $\varepsilon_{g_2}(C)$ is represented in figure 7 and the proposed two-qubit fault-tolerant error gate threshold is $\varepsilon_{thg_2}(C \to \infty) = 1.3 \ 10^{-5}$. This value could be considered as the strongest threshold obtained in this paper, permitting us to conjecture a full fault-tolerant quantum computation if $\varepsilon < \varepsilon_{thg_2}(C \to \infty)$ using noisy gates affected by a $0 < \gamma = \varepsilon/C$ error probability.

# 5. CONCLUSIONS

We have carried out a classical simulation using a classical computer, of a quantum error correction process applied periodically in time where the qubit is encoded by means of the [[7,1,3]] CSS quantum code (without concatenation) and the syndrome measurement is carried out using Shor's method. Evolution and gate errors reflecting decoherence and hardware errors are introduced into the recovery method. Decoherence is simulated via an isotropic depolarising channel error model.

In order to understand the encoded fidelity behaviour, the error equivalence is studied providing a simple scheme to describe the error effect on the codewords. The encoded fidelity is almost constant when the qubit coefficient changes, so the logical $|0_L\rangle$ state has been used in the numerical simulation. We have applied the decoherence error model to calculate the memory error threshold, considering an encoded qubit sent through a quantum channel affected by a memory error probability $\varepsilon$. If the condition $\varepsilon < \varepsilon_{mth}(C\rightarrow\infty) = 2.9 \cdot 10^{-5}$ is fulfilled, we can infer that there is a set of noisy gates with an error probability $0 < \gamma = \varepsilon/C$, able to construct a noisy fault-tolerant error correcting network stabilizing the qubit in the memory. In addition, this threshold assures a benefit of the encoded process compared to the non-encoded version.

The model also provides a one-qubit gate threshold. Given a non-encoded one-qubit gate, if the memory error probability is $\varepsilon$, and $\varepsilon < \varepsilon_{thg_1}(C \rightarrow \infty) = 2.7 \cdot 10^{-5}$, we can assume that the transversal encoded one-qubit gate version using the [[7,1,3]] code, is advantageous with respect to the non-encoded gate. The objectives reached are: it provides a smaller uncorrectable error probability than the non-encoded gate and it affords a fault-tolerant behaviour avoiding an excessive error accumulation.

Finally, a two-qubit gate error threshold has been estimated: $\varepsilon_{thg_2}(C \rightarrow \infty) = 1.3 \ 10^{-5}$. This value is the smallest threshold obtained in this paper, permitting us to expect a full fault-tolerant quantum computation if $\varepsilon < \varepsilon_{thg_2}(C \rightarrow \infty)$ and using noisy gates affected by a $0 < \gamma = \varepsilon/C$ error probability.

Obviously, these values depend on the encoding, the recovering process and the degree of parallelism in the network, related to the physical implementation of the quantum computer. Even though we expect that concatenation in the code would improve the recovery, the preliminary results obtained in this work allow us to be reasonably optimistic about the stabilization of the information in the memory as well as the possibility of carrying out a long enough computation to implement a quantum algorithm. We expect that better encoding and recovery methods would provide higher bounds for $\varepsilon$.

## ACKNOWLEDGEMENTS

This work has been supported by Spanish Ministry of Science and Technology Project BFM2002-01414.

17 In this section the X, Y and Z error operators do not include a hat in their notation in order to clarify the symbols.

FIGURES

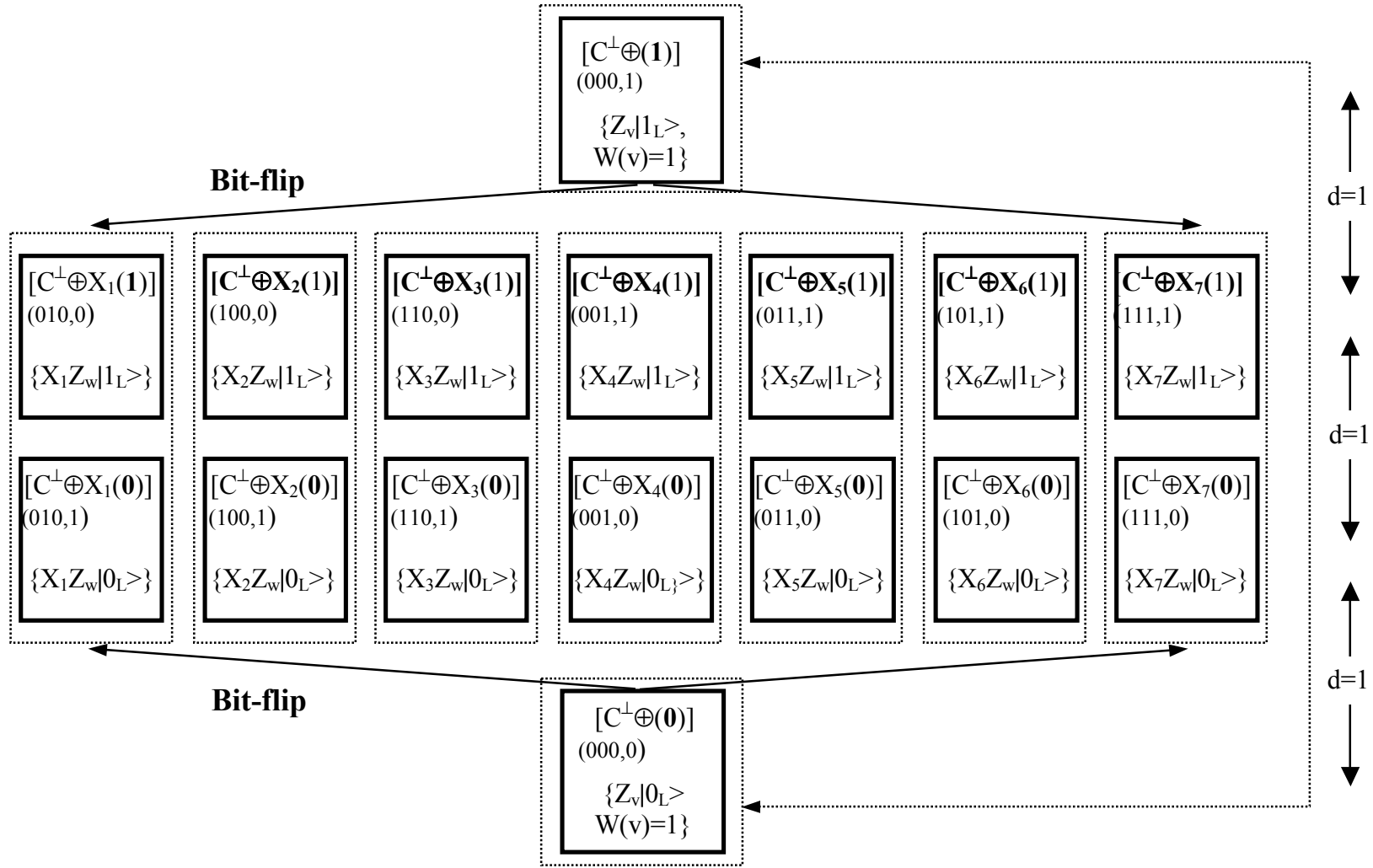

Figure 1: Relationship between the classical codewords and quantum states of a CSS [[7,1,3]] quantum code and the effect of bit-flip and phase-flip errors. Sixteen full line small boxes have two meanings. At the top of each box, the $C^\perp$ coset codewords notation appears as $[C^\perp \oplus X_v(\mathbf{a_L})]$ with (a = 0,1 and $W(\mathbf{v}) = 1$). Each coset includes 8 classical codewords corresponding to the same number of quantum states forming a Hilbert subspace of dimension 8 noted by means of brackets. Its basis appears at the bottom of each full line box as $\{X_v Z_w | \mathbf{a_L}\rangle\}$ with $W(\mathbf{v}) = W(\mathbf{w}) = 1$. A four component vector is shown below the coset notation as $(s_1, s_2, s_3; s_4)$. The first three components are the C = [7,4,3] syndrome and the fourth one completes the $C^\perp$ = [7,3,4] syndrome. Seven vertical dashed boxes represent C cosets, each of them including two $C^\perp$ cosets. The eighth C coset is constitute by the top and bottom boxes. On the right, the distance between the cosets illustrate the distance three of the [[7,1,3]] quantum code.

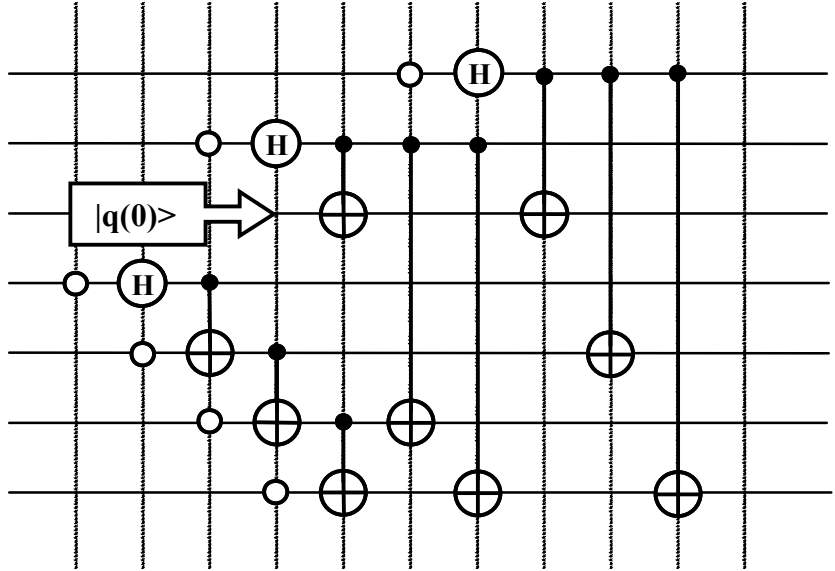

Figure 2: Network used to encode the initial qubit state |q(0)> = a|0> + b|1> into $|Q(0)_E\rangle = (a|0_L\rangle + b|1_L\rangle)$. The symbols are: H Hadamard gate, ○ is the $|0\rangle$ initial physical qubit and the other symbols are CNOT gates. Each vertical dotted line corresponds to a time step. The |q(0)> is introduced as the third initial qubit state.

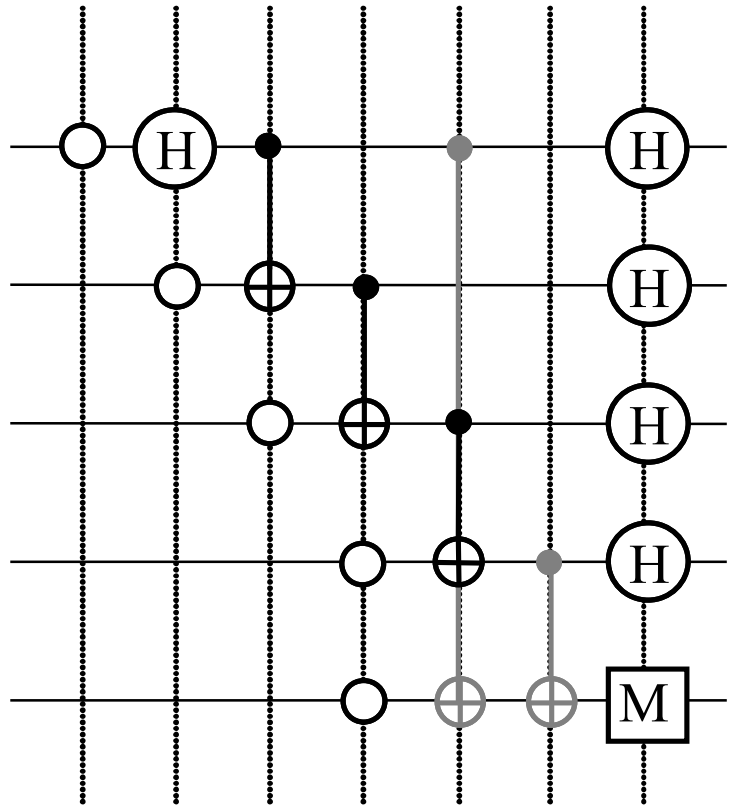

Figure 3: Network to prepare Shor's ancilla. The two grey CNOT gates make a bit-flip error checking. M represents a destructive measurement.

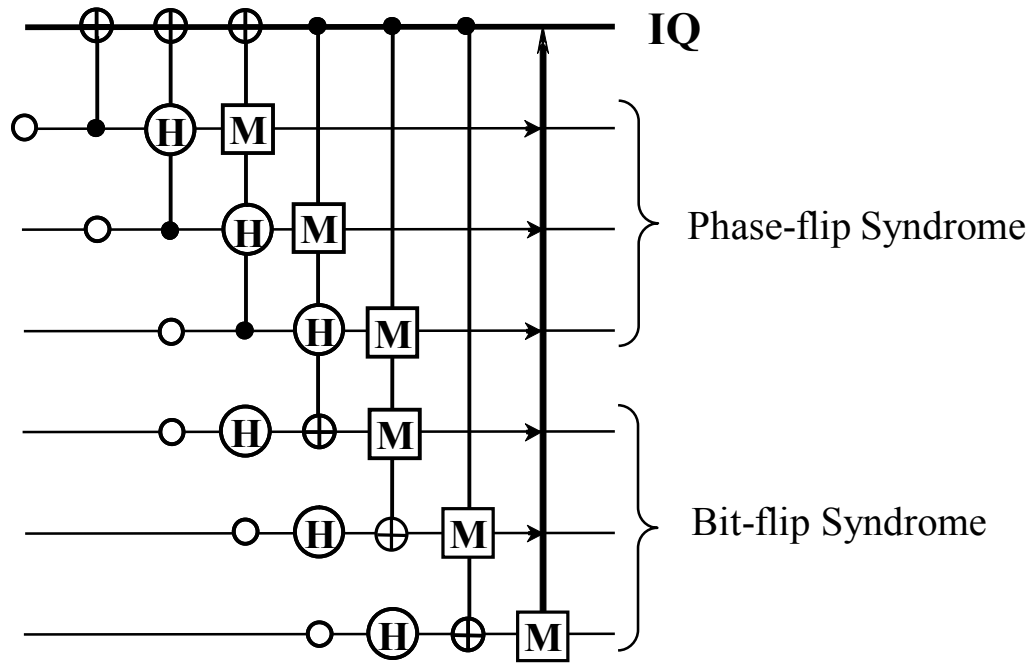

Figure 4: One syndrome measurement. Each initial ○ represents a 4-qubit cat state. M indicates its parity measurements providing one bit of syndrome. For a three fault-tolerant syndrome measurement, this network is repeated three times. After the last measurement, the IQ correction is carried out (vertical arrow).

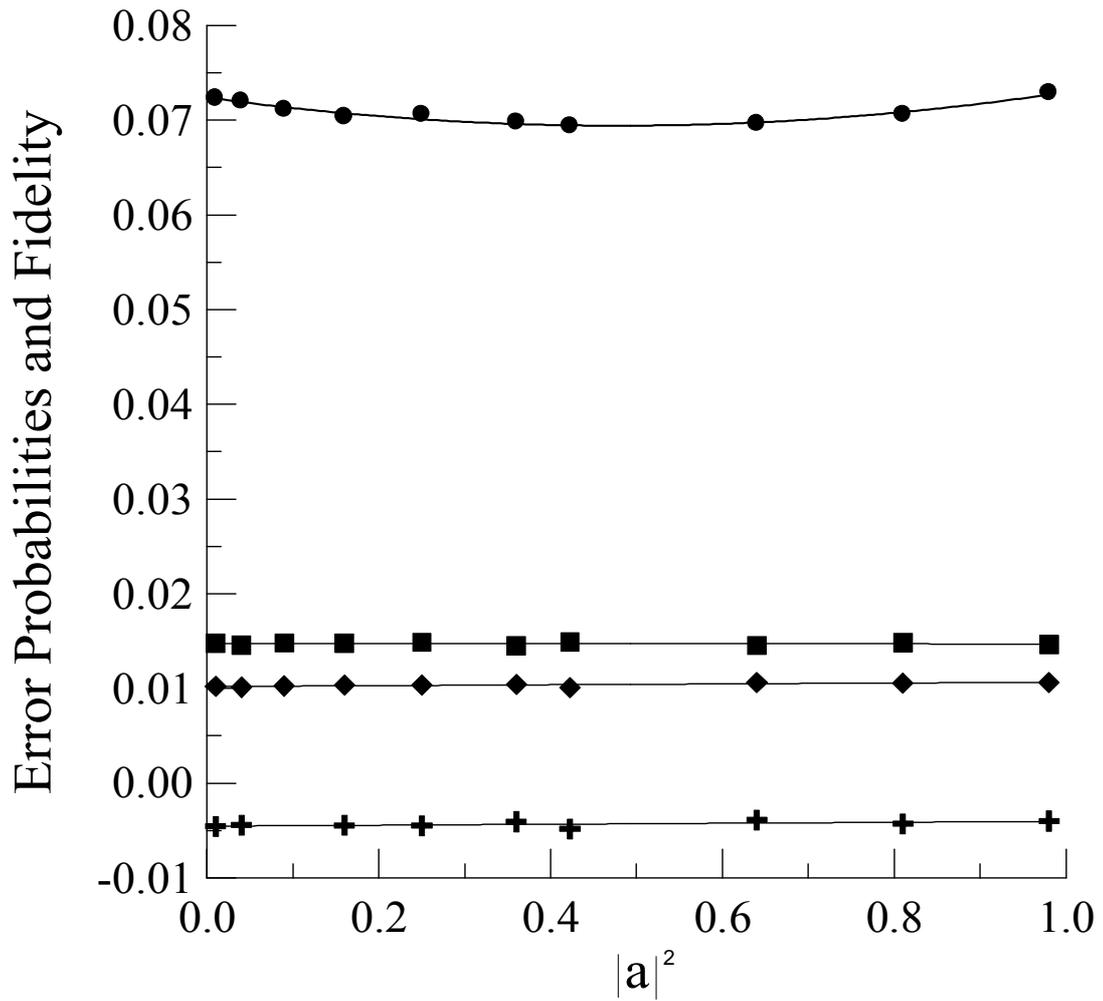

Figure 5: Simulation results for the fidelity and error probabilities vs. the a coefficient of the qubit, for $\varepsilon$(evolution error) = 0.002 and $\gamma$(gate error) = 0.02. Vertical axis represent: ● fidelity $F(\varepsilon,\gamma,t=20,a)$, ■ $\eta_3^p$, ♦ $\eta_3^b$ and + $\Delta\eta_3$. The last three probabilities do not depend on the a coefficient, having the slopes -6.08 $10^{-5}$, 4.6 $10^{-4}$ and 5.1 $10^{-4}$ respectively.

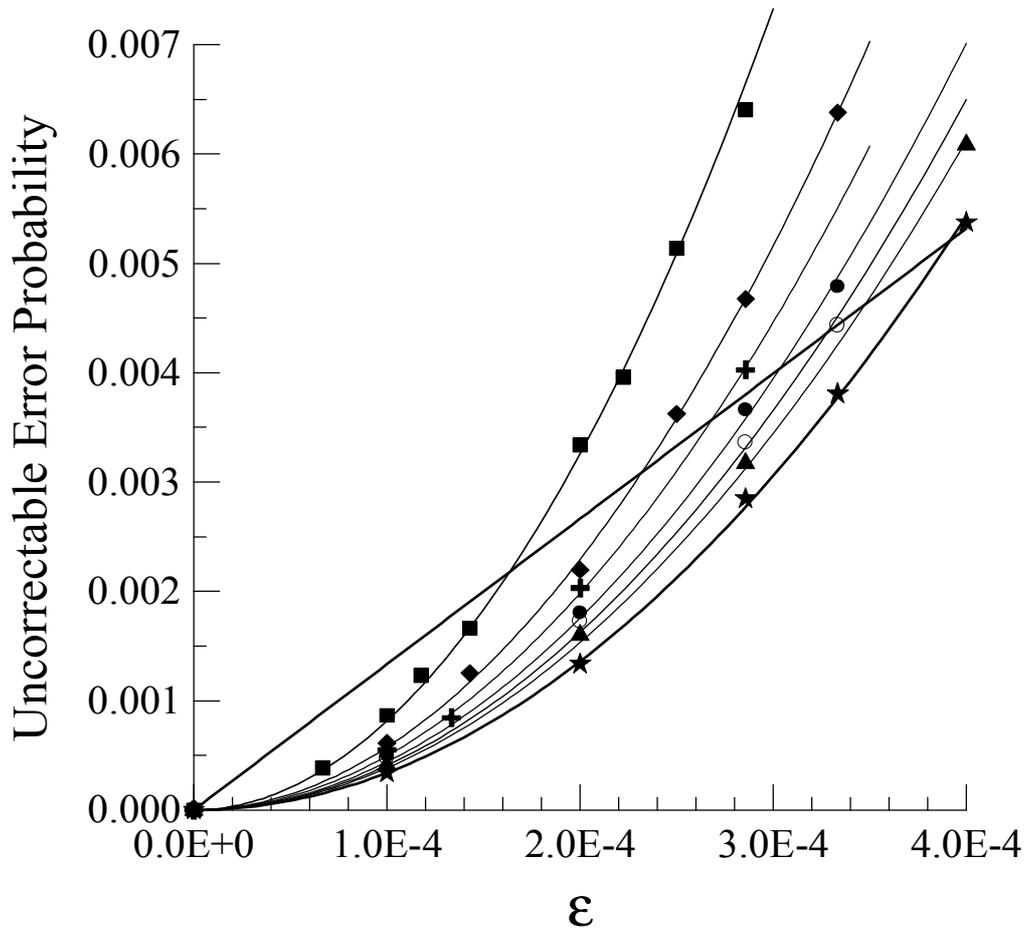

Figure 6: Comparison between the unrecoverable error probability for the non-encoded qubit after 20 time steps, $P_{NE} = 1-(1-2\varepsilon/3)^{20}$, and the same probability for the encoded and corrected qubit, $P_E(\varepsilon,C,t=20)$, vs. $\varepsilon$. The quadratic curves correspond to $P_E(\varepsilon,C,t=20)$ for different error relationships $C = \varepsilon/\gamma$: ■ 0.3, ◆ 0.5, ✚ 0.8, ● 1, ○ 1.5, ▲ 2 and ★ $\gamma=0$.

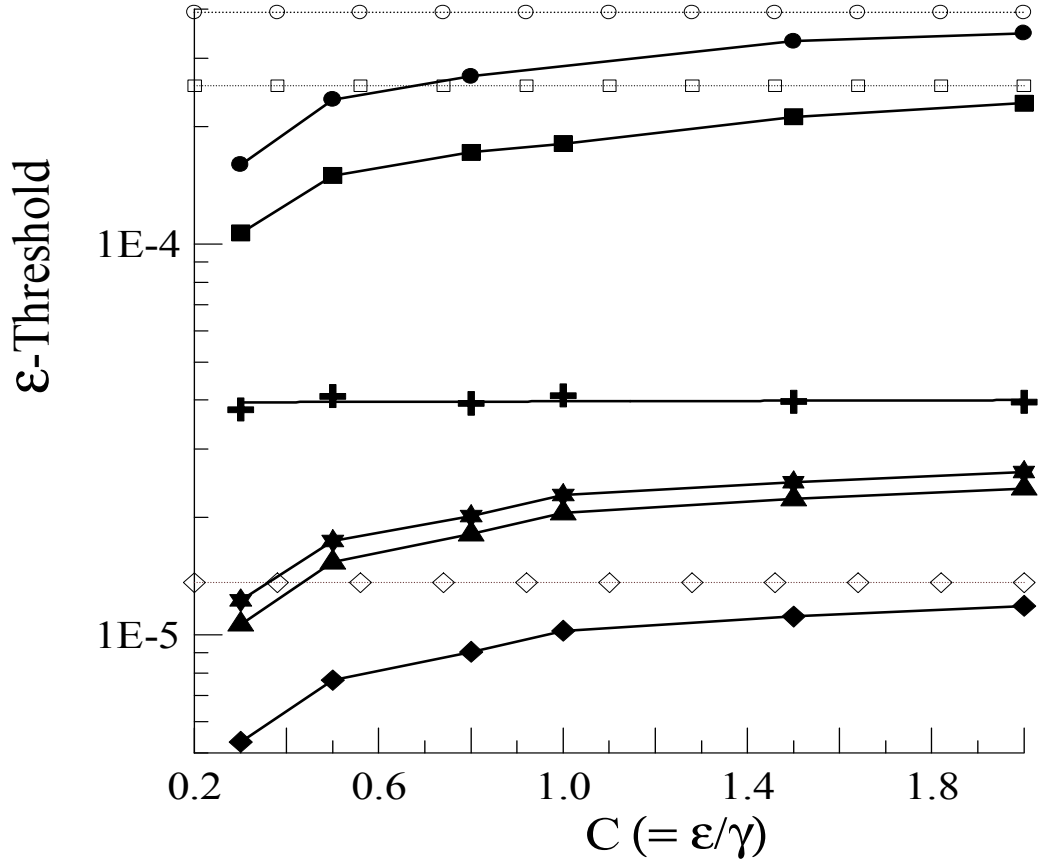

Figure 7: Thresholds curves vs. the $C = \varepsilon/\gamma$ relationship. Dashed lines with open symbols, represent the threshold limit for $C \to \infty$ of the curves with the same full symbol. Thresholds: ● $\varepsilon_{pth}(C)$, ■ $\varepsilon_{sth}(C)$, ★ $\varepsilon_{mth}(C)$, ✚ $\varepsilon_{g_1}(C)$, ▲ $\varepsilon_{thg_1}(C)$, ◆ $\varepsilon_{thg_2}(C)$.

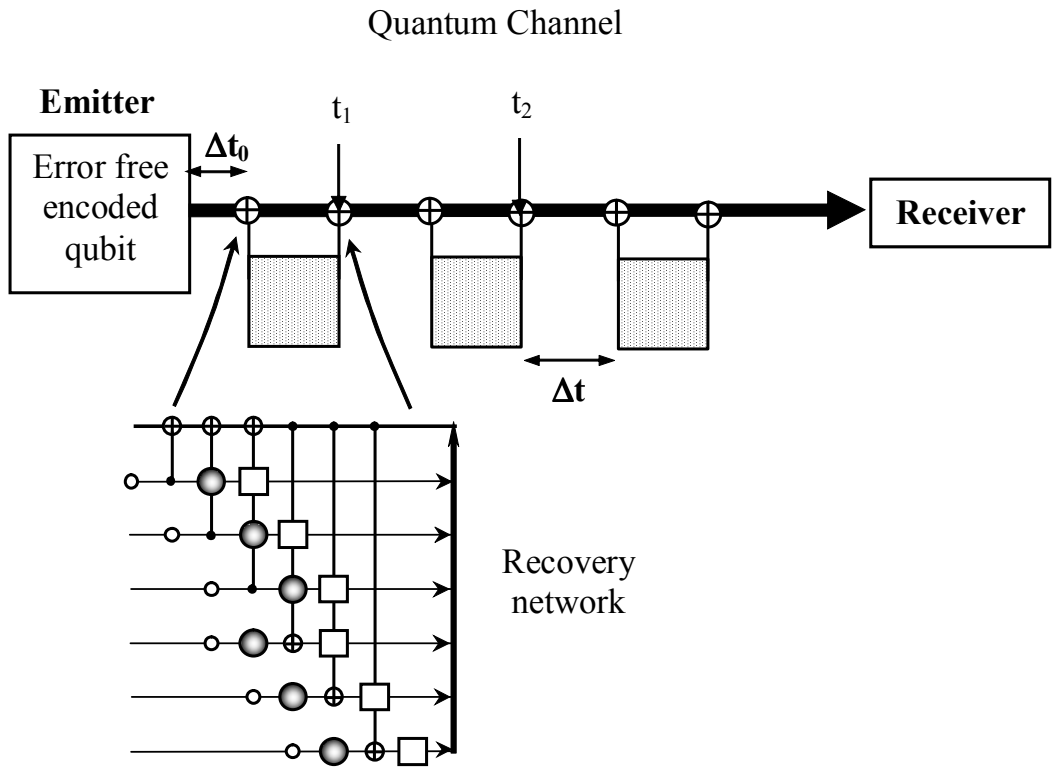

Figure 8: $|Q(0)_E\rangle$ qubit recovery through a noisy quantum channel. The first recovery is made after $\Delta t_0 =1$ time steps and $\Delta t=1$ is the time between consecutive corrections.

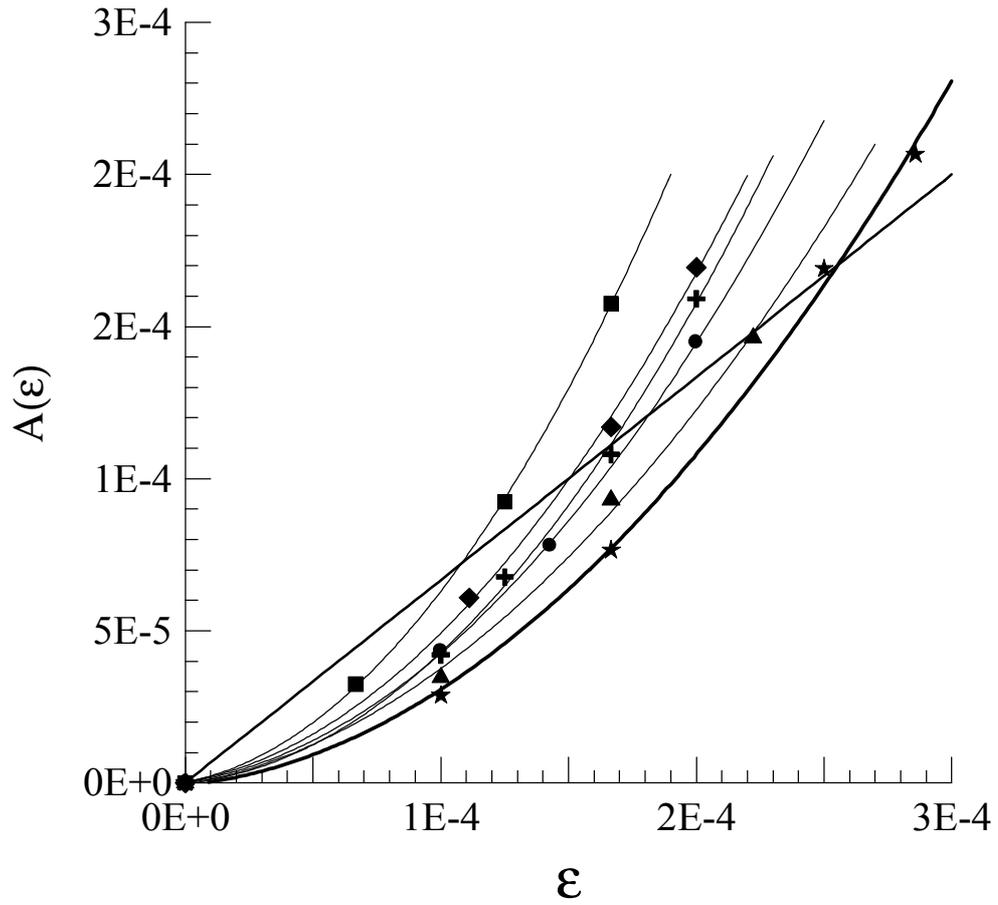

Figure 9: Slope (A(ε)) of the linear fitting of the encoded fidelity vs. the memory error ε probability. The straight continuous line represents the naked qubit fidelity slope $2\varepsilon/3$.

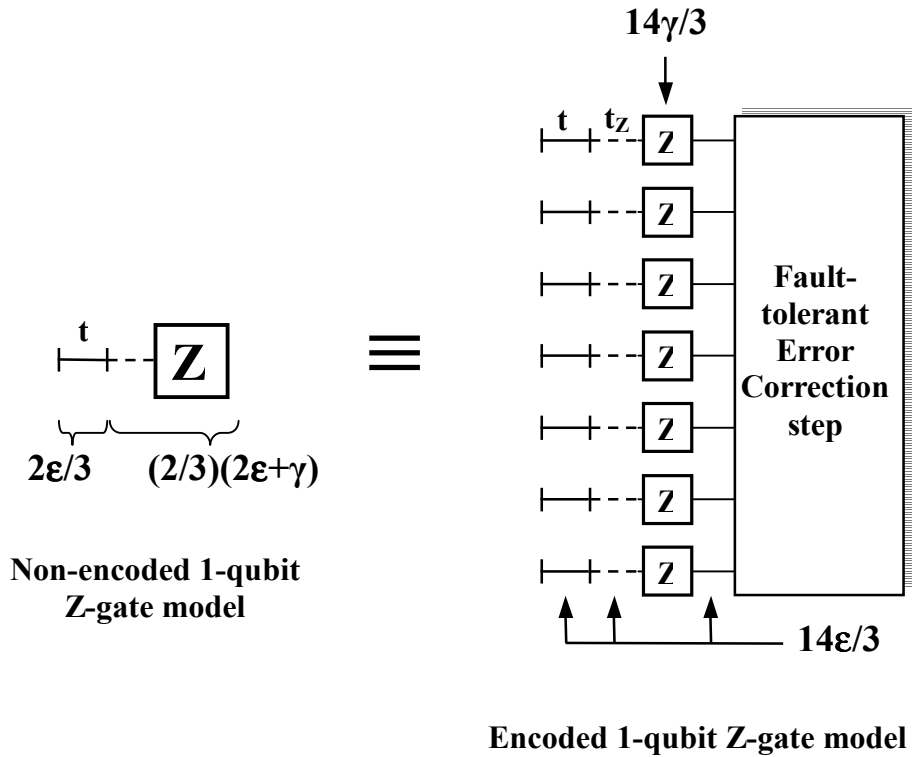

Figure 10. Simulation model used in the one-qubit gate threshold estimation. On the left, Z non-encoded gate. On the right transversal fault-tolerant version for the Z gate. Each one-qubit gate includes a previous time step t and an internal time step $t_Z$ (dashed line).

Table I

| C=ε/γ | $D_2$ | $D_1$ | $G_1$ |
|---|---|---|---|
| 0.3 | 81440.2 | 489. | 93900.2 |
| 0.5 | 57385. | 409.6 | 65195.7 |
| 0.8 | 49597.1 | 364.1 | 55228.7 |
| 1 | 43843.2 | 343.8 | 48749.7 |
| 1.5 | 40618 | 331.3 | 44814.5 |
| 2 | 38286.5 | 324.4 | 42135.7 |
| ∞ | 33961. | 290.8 | 36715.6 |